# Atomistic Mechanism of Phase Transition in Shock Compressed Gold Revealed by Deep Potential

Bo Chen[1], Qiyu Zeng[1], Han Wang[2], Shen Zhang[1], Dongdong Kang[1], Denghui Lu[3], Jiayu Dai[1, *]

[1]*Department of Physics, National University of Defense Technology, Changsha 410073, P. R. China*
[2]*Laboratory of Computational Physics, Institute of Applied Physics and Computational Mathematics, Beijing 100088, P. R. China*
[3]*HEDPS, CAPT, College of Engineering, Peking University, Beijing 100871, China*

A detailed understanding of the material response to rapid compression is challenging and demanding. For instance, the element gold under dynamic compression exhibits complex phase transformations where there exist some large discrepancies between experimental and theoretical studies. Here, we combined large-scale molecular dynamics simulations with a deep potential to elucidate the dynamic compression processes of gold from an atomic level. The potential is constructed by accurately reproducing the free energy surfaces of density-functional-theory calculations for gold, from ambient conditions to 15 500 K and 500 GPa. Within this framework, we extend the simulations up to 200 000 atoms size, and found a much lower pressure threshold for phase transitioning from face-centered cubic (FCC) to body-centered (BCC), as compared to previous calculations. Furthermore, the transition pressure is strongly dependent on the shock direction, namely 159 GPa for <100> orientation and 219 GPa for <110> orientation, respectively. Most importantly, the accurate atomistic perspective presents that the shocked BCC structure contains unique features of medium-range and short-range orders, which is named "disorders" here. We propose a model and demonstrate that the existence of "disorders" significantly reduces the Gibbs free energies of shocked structures, therefore leading to the lowering of the phase transition pressure. The present study provides a new path to understand the structure dynamics under extreme conditions.

The understanding of atomic structures under extreme conditions is of great importance in industrial applications and scientific studies, covering multidisciplinary fields of physics [1-3], chemistry [4, 5], materials [6], geophysics and astrophysics [7, 8]. Shock compression, served as a typical method to generate ultrahigh pressure-temperature (*P-T*) conditions [9, 10], leads to discoveries of unique thermodynamic states [11-13], unusual physical properties [14, 15], and unexpected meta-stable structures [7, 16-18]. With the development of *in-situ* time-resolved x-ray diffraction (XRD) methods, more attentions are paid to the phase transition process of shocked matters [17-21] rather than measuring the equation of state (EOS). Innumerous new phenomena have been observed and many experimental results present significantly different characteristics of structural transformation path between the dynamic and static compression, but the mechanism is far away to be known. There is a growing demand for theories from the atomic scale to understand the intrinsic mechanisms.

Gold (Au) is a typical close-packed face-centered-cubic (FCC) structure at ambient conditions, and it can maintain FCC structure under static pressure up to several hundreds of GPa [13, 22]. Thus, many high-pressure experiments based on diamond-anvil-cell (DAC) prefer utilizing Au as calibration material [23]. Under isothermal compression at room temperature, FCC gold undergoes the transitions to hexagonal close-packed (HCP) structure at 151 to 410 GPa in theoretical calculations [24-27], and at ~248 GPa in experiments [22]. *Ab initio* calculations predict that FCC to body-centered-cubic (BCC) transition would not occur until 230 GPa. For samples under dynamic compression, heating is inseparable to the compression, and the locus of possible shock states is a curve in the pressure-density-temperature space called the Hugoniot curve. In the phase diagram of Au calculated by Ref. [25], the Hugoniot curve directly crosses the melting curve instead of the FCC-BCC boundary. It means the gold only undergoes melting transition and no FCC-BCC transition would be observed under dynamic compression. Extraordinarily, due to the recent construction of high-quality *in situ* XRD tool, two phase transition points from FCC to BCC have been observed at 223 GPa [17] or 150~176 GPa [18] under two different shock-compression experiments. The FCC-BCC phase boundary is deemed to exist under much higher *P-T* conditions [25] in theoretical calculations, largely deviating from these experimental results. This controversy indicates that the dynamic process is significantly different between the static and shock compression, causing huge disparity of phase transition pressures. Although *ab initio* method can precisely predict the thermodynamic properties of shocked-compressed Au, it fails to capture the dynamic response process of the atoms and the evolution of microstructures under shock compression due to the limit of simulation sizes. In addition, molecular dynamics (MD) simulations based on semiclassical interatomic potentials is difficult to describe the interatomic interactions with so wide temperature and density range. It is urgently needed to obtain the detailed response of matters under rapid compression and reveal the intrinsic mechanisms in an accurate and large-scale way.

In this work, we performed MD simulations to investigate the dynamical structural transformation of shock compressed gold by utilizing a deep-neural-network potential. Our results show that the FCC Au can maintain its initial structure below around 38 GPa, and stacking faults with HCP structures begin to appear at higher impact pressures. The structural transformation path of Au is strongly dependent on the crystal orientation of shock wave propagation. According to the structure analysis, we find that the BCC structure generated by shock compression contains many locally ordered structures compared with the perfect crystals, and free-energy calculations prove that these specific local structures play key roles in lowering the FCC-BCC transition pressures. These results are in much good agreement with the recent experiments and can demonstrate the physics of their different phase boundaries. We provide an atomic insight into the dynamical phase transitions of compressed noble metals and bridge the existing knowledge gap in the mechanism of the FCC-BCC phase transition.

*Simulations of shock compression.* To mimic the propagation of shock waves, a simulation methodology combined with the Navier-Stokes equations for compressible flow and MD simulations has been performed for this study. All atoms in the system update positions and velocities following the modified Lagrangian, so as to restrain the systems to the Hugoniot-based thermodynamic conditions. The calculations are based on the method of Multi-Scale Shock Technique (MSST) [28] implemented under LAMMPS framework [29] (see Supplementary Material [30] for details). Prior to popular non-equilibrium molecular dynamics (NEMD) approaches, the MSST method is able to obtain the EOS and structures after shock compression without huge simulation systems (Fig. S1 [30]). Using classical interatomic potentials, the NEMD simulations with 8 million atoms have been complemented at three typical pressures (51.5, 159, 392 GPa), and the EOS, as well as structures, agrees well with MSST simulations (Fig. S2 [30]). Based on the recently developed deep potential for MD (DPMD) method [31, 32], a deep potential (DP) is generated in this work, which can accurately learn and predict the total energies, pressures, and forces at *ab initio* level. In order to train the potential at a wide range of thermodynamics conditions, a concurrent learning scheme, Deep Potential Generator [33, 34] (DP-GEN), has been adopted to smartly select samples for density functional theory (DFT) calculations and minimize the computational consumption [30]. The results of four kinds of embedded atom method (EAM) potentials [35-40] and on-the-fly DFT calculations (DFT-MSST) have been compared with the DP in terms of the EOS and structures. The DP matched well with experimental data and performs much more accurately in the predictions of shock compressed Au than all EAM potentials (Fig. S4 [30]). MSST simulations have been carried out for shock compression on the single crystal FCC gold along different orientations, <100> and <110> with the shock velocity of 4.0-8.0 km/s. A series of tests have been performed and 200 000 atoms are required to obtain a convergent results for shock wave propagation along <110>-direction (Fig. S5

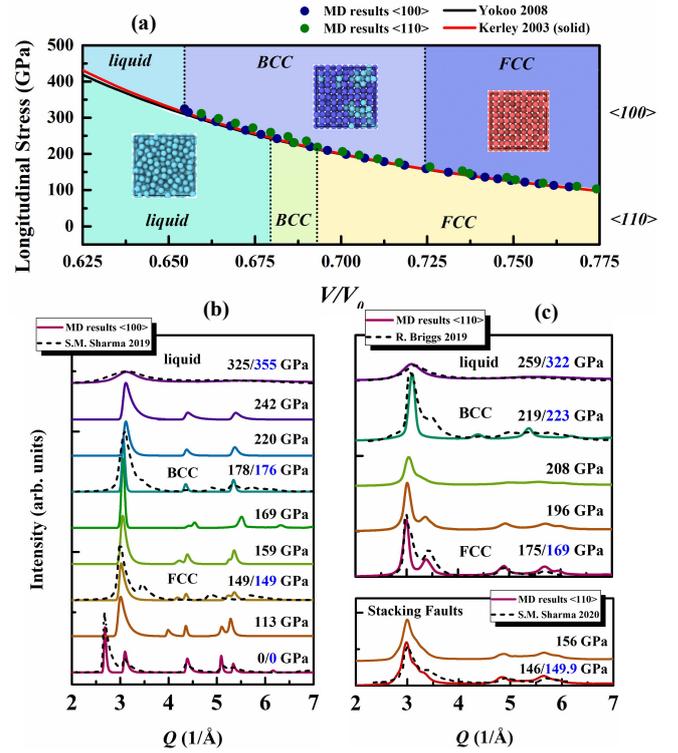

FIG. 1. (a) is the stress-volume states and structural transformation path of shock wave propagation through gold along <100> direction and <110> direction. (b) and (c) are the XRD profiles for shock compressed gold. Solid curves are the signals of structures obtained from MD simulations at different shock pressure. Black dash curves are the signals of structures from experimental results at shock pressures corresponding to the blue labels. The XRD intensity from experimental results is zoomed to a similar scale to the MD results.

[30]).

The pressure against volume compression for shocked Au is shown in Fig. 1a. The results show that the EOS is almost independent on the propagation direction of shock waves, which is in good agreement with the previous continuum measurements [13, 18, 41]. By computing the structure factor at each reciprocal lattice point (with a radiation wavelength of 0.5266 Å), the XRD profiles of structures after shock were calculated (method originally described in Ref. [42]). The Gaussian peak with an exponential tail (Exp-Gauss) was used to fit the XRD scatters. A series of XRD signals of shock compressed Au obtained from the experiments (black dash curves in Fig. 1) are determined as the benchmark. For <100>-direction shock, the first observation of BCC peak is found at 159 GPa, and the BCC signals remain at higher pressures. Above 325 GPa, only liquid signal is observed, indicating that Au completely transitioned to molten state. For <110>-direction shock, the pressure of the first observation of BCC peak increases to 219 GPa, and the melting pressure decreases to 259 GPa. According to the XRD signals, the entire Hugoniot curve can be divided into three regions according to FCC/BCC/liquid structures, represented by the areas with different filled colors in Fig. 1a. Compared with the shock along <100>-direction, there was a pressure delay for the FCC-BCC transition in shock compression along <110> direction, while the pressure range of BCC structure is reduced. After transition to BCC, it rapidly converted to molten state within tens of gigapascals. The same pressure delay of FCC-BCC transition was also observed in Au samples under shock compression along <100>-orientation rotated by ~1.14° (Fig. S7) [30]. It indicates that the atomic arrangement of the shock surface does not affect the *P-V* EOS, but significantly changes the structural transformation path of shock compressed Au. The onset pressure of the FCC-BCC transition is 159 GPa along <100> direction, while it is 219 GPa along <110> direction. These two transition points correspond to the results from two recent independent experiments, in which the phase transitions are found at 149~176 GPa in the work of Sharma *et al* [18], and 169~223 GPa in the work of Briggs *et al* [17], respectively. It should be noted that for <110>-direction at 146~156 GPa, the XRD profiles showed a broadened signal in FCC peak, which is the same as the structures under 149.9 GPa from the experimental results [21], in which this structure was determined as FCC with stacking faults in Ref. [21].

*Structure analysis from atomistic insight.* The XRD signal is considered as a long-range diagnosis method with average in space and time (the time resolution of XRD experiments is about 100 picoseconds), which cannot clearly describe the detailed microscopic structures in short or medium range and related dynamics. Therefore, the structure analysis was conducted from an atomistic insight. The adaptive common neighbor analysis (a-CNA) [43] has been utilized to classify atoms in crystalline systems through OVITO software [44], and the fraction of atoms identified in different types are shown in Fig. 2(a) and Fig.2(b). It is found that the onset pressure of phase transition was 113 GPa for <100>-direction, but through XRD diagnosis method the FCC-BCC transition pressure was determined as 159 GPa (marked with dash dot line in Fig. 2a and Fig.2b). It indicates the phase transition starts from some parts of structures and gradually progresses to long-range characteristics of BCC structure. For <110>-direction shock, the transition pressure through a-CNA is in agreement with XRD analysis (219 GPa). However, the fraction of BCC structures is less than 50% indeed at 219 GPa with a-CNA. Furthermore, in the shaded region (146~156 GPa), the structures are considered as stacking faults deduced from XRD signals in Ref. [21], and it is now confirmed as the coexistence structure of FCC-BCC-HCP through a-CNA.

It should be noted that in each case there are large number of atoms that do not show a certain crystal structure feature, which are identified as "Other" in a-CNA. It is because this method utilizes instantaneous structure to analyze the neighbor environment. If some atoms deviate from the equilibrium positions of the crystalline lattice, the method is unable to capture the structure characteristics. To give a view of the unrecognized atoms, the effective coordination number (ECN) model is introduced to determine the symmetry structures where a specific atom is surrounded by its neighbors at different distances [3, 30]. The ECN concept can be independent of the choice of the cutoff of bond length, and the temporal series of structures can be used to obtain time-averaged information and to analyze medium or short-ordered structures.

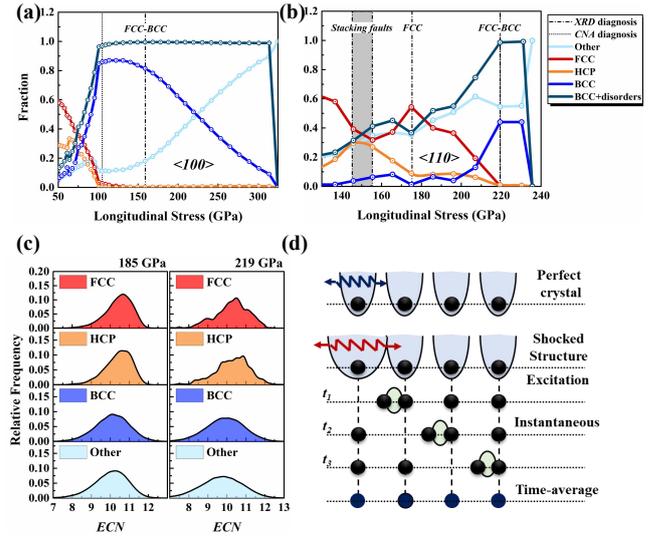

FIG. 2 The fraction of atoms with different structural features identified by a-CNA method as a function of longitudinal stress for shock compression along <100> direction (a) and <110> direction (b). Dash dot lines and dot lines highlight the pressures by utilizing XRD and a-CNA diagnosis method for specific structures of FCC-BCC transition point and stacking faults. (c) The ECN distribution of atoms with different structure types with shock wave propagating along <110> direction. (d) The schematic diagram of the BCC with disorders after shock compression (Detailed description of the model see Section 8 in the Supplementary Material [30]).

We show the ECN of Au structures on shock compression along <110> direction in Fig. 2(c). Ten snapshots were collected from MD simulations during 100 ps. As shown in Fig. 2(c), the probability distribution of ECN has been presented in each type of atoms identified through a-CNA. The distribution of BCC atoms is apparently different from HCP and FCC atoms. We select the structures at two shock pressure of 185 GPa, and 219 GPa to investigate the change of ECN distributions at different thermodynamics conditions, and the distribution functions have a trend of broadening with the increasing shock pressures (*i.e.,* temperature) due to more intensive thermal motion of atoms. The ECN distributions of "Other" type atoms are consistent to those of BCC structures, even when the distribution changes under different conditions. These observations are also found at other shock velocities (Fig. S8 [30]). It indicates that "Other"-type atoms can be charactered as BCC structure in the long-time statistical average. To explore the feature of the specific structures, we have compared the radial distribution functions (RDF) of FCC Au, shock-induced BCC Au, and perfect BCC crystal Au at the same thermodynamic conditions (Fig. S9 [30]), and the RDF suggests that more short-ordered or medium-ordered structural features emerge after shock compression. Combing with above diagnosis method, it can be concluded that the shock-induced BCC structures have two kinds of characteristics, named "instantaneous disordered structure" (others) and "time-averaged crystal". A physical model is proposed to depicted the process, and the schematic diagram is shown in Fig. 2(d). We considered the crystal as a collection of atoms with local oscillations. After shock compression with rapid and strong uniaxial strain, it will cause more intensive atomic oscillations than that under static compression. This process will stimulate some atoms to deviate further away from the equilibrium positions of the crystal lattice in local area. During the time evolution, these atoms exhibit as unrecognized structures from instantaneous insight, while the equilibrium positions of these atoms under the long-time statistical average perform the feature of perfect crystals. The intensive vibration will also cause the excitation of collective motions, which can be charactered as excess quasi-particles (similar to phonons) in the systems, named as "disorders" here. Thus, we recognize the fraction of "Other" into BCC atoms, and redefine the evolution of BCC with disorders as the steel blue curves shown in Fig. 2(b). The Gibbs free energy of the hybrid structures can be decomposed into the free energy of perfect BCC lattices and the Gibbs free energy of disorders:

$$G = G_{BCC} + G_{disorders}$$
$$\phantom{G} = (H_{BCC} + H_{disorders}) - T(S_{BCC} + S_{disorders}) \quad (4)$$

Here $G$ is the Gibbs free energy, $H$ is the enthalpy and $S$ is the entropy of the system, respectively. According to the formula, the presence of disorders will both increase the entropy and enthalpy of the systems, so that the free energy should be practically calculated to determine the effect of the disorders on the phase boundary.

*New Phase diagram of shock-compressed Au.* We have calculated the Gibbs free energy difference between FCC Au and two kinds of BCC structures of perfect BCC structure (without disorders) and shock-induced BCC structure (with disorders). The shock-induced BCC structure was generated from MSST simulations in the shock velocity of 5.5 km/s (<100>-direction) and 6.1 km/s (<110>-direction), corresponding to the phase transition pressures of 159 and 219 GPa respectively. To avoid the occurrence of structural transformation, we restricted the free energy calculations of shock-induced BCC structure around the Hugoniot curves. The FCC and perfect BCC structures were directly constructed and equilibrated at the same thermodynamic conditions. The non-equilibrium processes based on the thermodynamic integration (neTI) method [45, 46] have been performed for the free energy calculations. Instead of constructing a series of equilibrium states on the pathway between two thermodynamic states of interest, this approach considers the path in terms of an explicitly time-dependent processes. Besides, a reversible-scaling (RS) method [47] is combined for calculating the free energies at different temperatures in a wide range of *P-T* efficiently (details in Supplementary Material [30]).

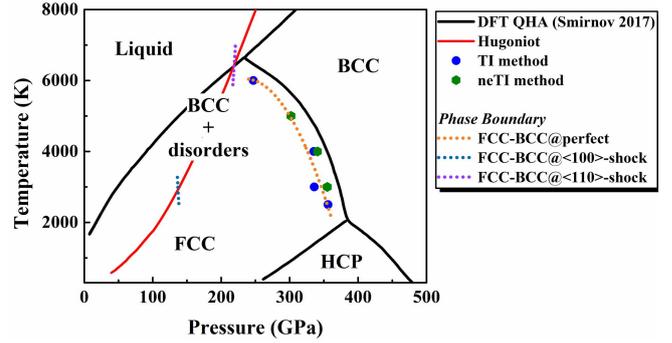

FIG. 3 The phase diagram of shock-compressed Au. The red line is the Hugoniot curve of shock-compressed Au. Black solid lines are the phase boundary obtained from *ab initio* calculations of Smirnov et al by using quasi-harmonic approximation (QHA) method. Scatters are the thermodynamic points where $\Delta G = 0$. The phase boundary is obtained by cubic polynomial fitting, represented as dot lines.

Regarding to the structures of BCC with disorders in shock compression, we refined the phase diagram of Au based on the results obtained from *ab initio* calculations [25], which were obtained by using quasi-harmonic approximation (QHA) method. For the Gibbs free energy between perfect BCC and FCC, the phase boundary calculated by neTI method located at the pressure region from 280 GPa to 350 GPa. It is close to the prediction of QHA results with a shift to lower pressure, because thermodynamic integration (TI) method includes the anharmonic effect on the free energy landscape. We have also appended the standard equilibrium TI method [48] to calculate the phase boundary as a benchmark. It is in agreement with calculations from the non-equilibrium method, and the subtle difference is due to the size variance of these two methods (see Section 9 in the Supplementary Material [30]). Most importantly, for free energy difference between FCC and BCC with disorders, the phase boundary is much lower than that of perfect BCC. The transition from FCC to the structure of BCC with disorders can take place almost at the pressure above 140 GPa (<100> direction) and above 220 GPa (<110> direction), corresponding to the onset pressure of FCC-BCC phase transition calculated in MSST simulations. It is clearly shown that the formation of disorders (short- or medium-ordered structures) induced by dynamical shock compression plays a key role in lowering the phase transition pressure from FCC to BCC. This physical model can be able to apply in other systems where the phase transition pressure is lowered due to shock compression [19]. In addition, by using EAM potentials of Sheng and Starikov, we have obtained the similar evolution of atoms with structural features, but the transition pressures and microstructures are far away from the experiments.

In conclusion, considering the dynamic behaviors of microscopic structures under shock compression, this work has explored the structural transformations path of FCC gold under shock compression in a wide range of *P-T* conditions by using the deep potential. A series of characteristic structures have been observed from the emergence of stacking faults to completely melting up to 325 GPa. Through XRD diagnosis method, an orientation-dependent structural transformation path is observed, and much lower FCC-BCC transition pressure is found comparing with previous calculations. It is much consistent with recent experiments. Combing with a-CNA and ECN model, the microscopic pictures of the generated BCC structures by shock compression have been demonstrated, which exhibits the characteristics of "instantaneous disordered structure" and "time-averaged crystal". We propose a new model to depict the physical diagram that the shock compression will enlarge the atomic oscillations, which can be considered as the excitation process to produce many quasi-particles, named as "disorders". These disorders will reduce the Gibbs free energies of perfect BCC crystal, consequently largely lowering the transition pressure. The microscopic details highlight the vital role of the different physical characteristics between dynamic and static compressions, well explaining the contradiction between recent experiments and prior theoretical calculations. By this way, we can better understand what has happened in the dynamic compression processes and what determines the phase transitions from atomic scale, giving a new path to dig out the deep mechanism regarding to this sort of experiments.

We are grateful for the insightful discussions with Prof. Andrew Ng, Prof. Jianmin Yuan and Dr. Mianzhen Mo. This work was supported by the National Key R&D Program of China under Grant No. 2017YFA0403200, the Science Challenge Project under Grant No. TZ2016001, the National Natural Science Foundation of China under Grant Nos. 11774429 and 11874424, and 12047561, the NSAF under Grant No. U1830206, the China Postdoctoral Science Foundation No. 2019M664024. HW is supported by the National Science Foundation of China under Grant No.11871110 and Beijing Academy of Artificial Intelligence (BAAI). All calculations were carried out at the Research Center of Supercomputing Application at NUDT.

*Corresponding authors: jydai@nudt.edu.cn